\pdfoutput=1
\documentclass[conference]{IEEEtran}
\IEEEoverridecommandlockouts
\usepackage{amsmath,stfloats}
\usepackage{graphicx}
\include{psfig}
\usepackage{epsfig}
\usepackage{array}
\usepackage{psfrag}
\usepackage{amssymb}
\usepackage{mathdots}
\include{psfig}
\usepackage{color}
\usepackage{pstricks,pst-node,pst-text,pst-3d,pst-plot}
\usepackage{psfrag}

\begin{document}
\title{Optimal Beamforming for MIMO Shared Relaying in Downlink Cellular Networks with ARQ}
\author{\large Ahmed Raafat Hosny $^\dagger$, Ramy Abdallah Tannious$^\ddagger$,
Amr El-Keyi$^\dagger$ \\ [.1in]
\small \begin{tabular}{c} $^\dagger$Wireless Intelligent Networks Center (WINC), Nile University, Smart Village, Egypt.\\
$^\ddagger$Aviat Networks, Santa Clara, CA, USA. \\
\small email: ahmed.raafat@nileu.edu.eg, ramy@alumnimail.utdallas.edu, aelkeyi@nileu.edu.eg \\
\end{tabular} }

\maketitle
\begin{abstract}
In this paper, we study the performance of the downlink of a
cellular network with automatic repeat-request (ARQ) and a half
duplex decode-and-forward shared relay. In this system, two
multiple-input-multiple-output (MIMO) base stations serve two
single antenna users. A MIMO shared relay retransmits the lost
packets to the target users. First, we study the system with
direct retransmission from the base station and derive a closed
form expression for the outage probability of the system. We show
that the direct retransmission can overcome the fading, however,
it cannot overcome the interference. After that, we invoke the shared relay and design the
relay beamforming matrices such that the
signal-to-interference-and-noise ratio (SINR) is improved at the
users subject to power constraints on the relay. In the case when
the transmission of only one user fails, we derive a closed form
solution for the relay beamformers. On the other hand when both
transmissions fail, we pose the beamforming problem as a sequence
of non-convex feasibility problems. We use semidefinite relaxation
(SDR) to convert each feasibility problem into a convex
optimization problem. We ensure a rank one solution, and hence,
there is no loss of optimality in SDR. Simulation results are
presented showing the superior performance of the proposed relay
beamforming strategy compared to direct ARQ system in terms of the
outage probability.
\end{abstract}
\section{INTRODUCTION}
Fading and interference are fundamental phenomena of wireless communications.
They limit the performance of many wireless systems.
Automatic-repeat-request (ARQ) protocols have been developed
as a mechanism to combat fading
and ensure reliable data transmission for wireless communication
systems. The idea of ARQ protocols is that a user
requests retransmission when a message is not correctly decoded. A
review of ARQ protocols is presented in \cite{wicker1995error} and
the performance of different ARQ protocols is discussed in
\cite{arqe}. In ARQ systems, there is a natural tradeoff between
the data transmission rate and ARQ re-transmission rate. In
\cite{wu2011coding}, it was shown that the performance of ARQ
systems can be improved by selecting the data transmission rate
based on the fading statistics. However, direct ARQ cannot
significantly improve communication reliability in multiuser
communication as interference is the main causes of transmission
failure.

This work proposes to manage the inter-cell interference of cellular networks
with ARQ using multiple-input-multiple-output (MIMO) shared relay.
Many techniques have been proposed to manage inter-cell
interference in cellular networks. The coordination of base stations (BSs)
can significantly improve the system performance. However, it
requires sharing large amounts of data between different BSs under tight coordination\cite{peters2009relay}.
Also, interference alignment and
cooperative systems require the availability of complete channel
state information at all the transmitting nodes. In contrast,
cellular relay networks present a promising practical solution not
only for improving the network coverage but also for managing
interference\cite{ramy2012}.
In \cite{practrelay}, a hybrid-ARQ (HARQ)
technique was proposed for wireless relay networks where the relay
helps the source to communicate with a single user terminal by retransmitting failed packets.
In \cite{relay_cop}, a comparison is presented
between the performance of decode-and-forward (DF) ARQ relay
systems relative to direct ARQ systems (where the retransmission
is performed by the BS) in terms of the outage
probability showing that the system performance can be
significantly improved with relaying. In
\cite{zimmermann2005performance}, the performance of distributed
HARQ protocol was shown to be much better than direct transmission
in terms of outage probability as the outage event occurs when the
relay and the destination fail to decode simultaneously. However,
in all this prior work, only single user ARQ systems were
investigated and the effect of multiuser interference was not
considered.

In this paper, we consider the downlink of a multiple-input-single-output (MISO)
cellular ARQ system where the BSs are equipped with multiple antennas. We
derive closed-form expressions for the outage probability in the
presence and absence of inter-cell interference showing that direct
ARQ cannot improve the outage probability of the system. Next, we
consider a cellular relay network where a MIMO half-duplex DF shared relay
retransmits the lost packets to the target users. We design the relay beamforming matrices such that
the signal-to-interference-and-noise ratio (SINR) is improved at
the users subject to power constraints on the relay. We consider
two cases for the relay retransmission. The first case is when
only one user fails where we derive a closed-form solution for the
relay beamformers. On the other hand when both transmissions fail,
we pose the beamforming problem as a sequence of non-convex
feasibility problems. We use semidefinite relaxation (SDR) to
convert each feasibility problem to a convex optimization problem.
We ensure a rank one solution, and hence, there is no loss of
optimality in SDR. Numerical simulations are presented showing
that the proposed beamforming algorithms can significantly improve
the outage probability compared to direct ARQ systems.
%
\section{miso cellular system with direct ARQ}
We consider the downlink of a cellular system containing two
BSs each serving a user terminal. For the sake of
simplicity, we consider a two-cell case where two BSs
serve two users. However, the proposed algorithms can be directly
extended to multiple cells. We assume that the transmitting BSs do
not have any any information about the channel coefficients of the links to the users.
Each BS is equipped with $N$ antennas while each user has
one antenna only. The $i$th BS transmits the data vector
$\boldsymbol{x}_{i}$ to the $i$th user where $\boldsymbol{x}_{i}$
has zero-mean with a covariance matrix $\boldsymbol{I}_{N}$ and
$\boldsymbol{I}_{N}$ denotes the $N\times N$ identity matrix. The
received signal at the $i$th user is given by
\begin{equation}
{{y}}_{i}=\sqrt{\dfrac{P}{N}} \sum_{j=1}^2
\boldsymbol{{h}}_{i,j}^H \boldsymbol{{x}}_{j} +{n}_{i}
\label{eq1_MISO}
\end{equation}
where $\boldsymbol{h}_{i,j}$ is the $N\times 1$ vector containing
the complex conjugate of the coefficients of the channel from the
$j$th BS to the $i$th user $i$.  All channel
coefficients are independent and identically distributed
circularly symmetric complex Gaussian random variable with zero-mean.
We assume that the variance of the direct channels, i.e., from
the $i$th BS to the $i$th user, is given by $\sigma_1^2$
and the variance of the interference channels is given by
$\sigma_2^2$. In \eqref{eq1_MISO}, ${n}_{i}$ is the zero-mean
circular Gaussian complex noise received at the $i$th user with
variance $\sigma^2$ and $P$ is the transmitted power from each
BS.

Let us consider the case where there is no interference,
i.e., $\boldsymbol{h}_{1,2}=\boldsymbol{h}_{2,1}=\boldsymbol{0}$.
In this case,  the mutual information between the $i$th BS and its corresponding user is given by
\begin{equation}
{I}^{\text{(SU)}}_{i}={\log}_{2}\left(1+\frac{P}{N\sigma^{2}}\boldsymbol{h}_{i,i}^{H}
\boldsymbol{h}_{i,i} \right) \label{eq24}
\end{equation}
The corresponding outage probability is given by
\begin{equation}
\begin{aligned}
P^{\text{(SU)}}_{\text{out},i}&=\text{Pr}\left\{{I}^{\text{(SU)}}_{i}<R
\right\}\\
&=\text{Pr}\left\{\tilde{\boldsymbol{h}}_{i,i}^{H}\tilde{\boldsymbol{h}}_{i,i}
<\frac{N\sigma^{2}}{P\sigma^{2}_{1}}\left(2^{R}-1\right) \right\}.
\end{aligned}
\label{eq25}
\end{equation}
where
$\tilde{\boldsymbol{h}}_{i,i}={\boldsymbol{h}}_{i,i}/\sigma_1$ and
$R$ is the attempted transmission rate by the BS. Since
$\tilde{\boldsymbol{h}}_{i,i}\sim
\mathcal{C}\mathcal{N}(0,{\boldsymbol{I}}_{N})$, then
$\tilde{\boldsymbol{h}}_{i,i}^{H}\tilde{\boldsymbol{h}}_{i,i}\sim
\mathcal{X}_{(2N)}^{2}$, i.e., Chi-square distribution with $2N$ degrees of
freedom. The outage probability (\ref{eq25}) is given by
\begin{equation}
P^{\text{(SU)}}_{\text{out},i}=Q_{\mathcal{X}_{(2N)}^{2}}\left(\frac{2N\sigma^{2}}{P\sigma^{2}_{1}}\left(2^{R}-1\right)
\right)
\end{equation}
where $Q_{\mathcal{X}_{(2N)}^{2}}\left(x\right)$ is the cumulative
density function of a Chi-square distribution with $2N$ degrees of
freedom.

Next, we consider the case with inter-cell interference.  The
mutual information between the first user and its BS in this case
is given by
\begin{equation}
{I}_{1}={\log}_{2}\left(1+
\frac{\frac{P}{N}{\boldsymbol{h}}_{1,1}^{H}{\boldsymbol{h}}_{1,1}}{\frac{P}{N}{\boldsymbol{h}}_{1,2}^{H}{\boldsymbol{h}}_{1,2}+\sigma^{2}}
\right) \label{eq26}
\end{equation}
The probability of outage of the first user is given by
\begin{eqnarray}
P_{\text{out},1}&\!\!\!\!=\!\!\!\!&\text{Pr}\left\{
\frac{{\boldsymbol{h}}_{1,1}^{H}{\boldsymbol{h}}_{1,1}}{{\boldsymbol{h}}_{1,2}^{H}{\boldsymbol{h}}_{1,2}+\frac{N\sigma^{2}}{P}}<
\gamma \right\}\\
&\!\!\!\!=\!\!\!\!& \text{Pr}\left\{ \sum_{k=1}^N y_k \!<\!
\frac{\gamma N\sigma^{2}}{P} \right\}
\end{eqnarray}
where $\gamma=2^{R}-1$, $y_k= \left|{{h}}_{1,1}(k)\right|^2 -
\gamma \left|{{h}}_{1,2}(k)\right|^2$, and ${{h}}_{i,j}(k)$ is the
$k$th element of the vector $\boldsymbol{h}_{i,j}$. Since
$\boldsymbol{h}_{1,1}\sim
\mathcal{CN}\left(\boldsymbol{0},\sigma_1^2 \boldsymbol{I}\right)$
and $\boldsymbol{h}_{1,2}\sim
\mathcal{CN}\left(\boldsymbol{0},\sigma_2^2
\boldsymbol{I}\right)$, then $y_k$ are independent identically
distributed random variables with probability density function
\begin{equation}
f_{y_k}\left(y_k\right) = \begin{cases} \frac{\lambda}{\lambda+\mu}\mu e^{-\mu y_k} &\mbox{if } y_k>0. \\
\frac{\mu}{\lambda+\mu}\lambda e^{\lambda y_k} & \mbox{if } y_k <
0. \end{cases} \label{eq31}
\end{equation}
where $\lambda=\frac{1}{\sigma_{1}^{2}}$ and
$\mu=\frac{1}{\gamma\sigma_{2}^{2}}$. The characteristic function
$\phi_{Y_k}(t)$ of the random variable $Y_k$ is given by
\begin{eqnarray}
\phi_{Y_k}(t)&\!\!\!=\!\!\!&\text{E}\left\{e^{j y_k t}\right\}\\
&\!\!\!=\!\!\!&\left(\frac{\lambda\mu}{\lambda+\mu}\right)
\left(\frac{1}{\lambda+jt}+\frac{1}{\mu-jt}\right) \label{eq32}
\end{eqnarray}
where $j=\sqrt{-1}$. Since the random variables $Y_k$ are
independent, the characteristic function of the random variable
$Z=\displaystyle \sum_{k=1}^N Y_k$ is given by \vspace{-2mm}
\begin{eqnarray}
\phi_{Z}(t)&\!\!\!=\!\!\!&\prod_{k=1}^{N}\phi_{Y_k}(t)\\
&\!\!\!=\!\!\!&\left(\frac{\lambda\mu}{\lambda+\mu}\right)^{N}
\left(\frac{1}{\lambda+jt}+\frac{1}{\mu-jt}\right)^{N}
\label{eq33}
\end{eqnarray}
The probability density function of $Z$ is the inverse Fourier
transform of $\phi_{Z}(t)$
\begin{equation}
f_{Z}\left(z\right)=\frac{1}{2\pi}\int_{-\infty}^{\infty}
e^{-jtz}\phi_{Z}(t)\, dt \label{eq34}
\end{equation}
The integration (\ref{eq34}) can be solved by using contour
integration. The result of the integration equals $2\pi \sum_{i}
r_i(z)$ where $r_i(z)$ is the residue of $e^{-jtz}\phi_{Z}(t)$ at
the $i$th singular point. We have two singular points
$t_{1}=j\lambda$ and $t_{2}=-j\mu$ repeated $N$ times. If the
singular point is repeated $N$ times, its residue is given by
\begin{equation}
r_{i}(z)=\lim_{{t}\to{t_{i}}}\frac{1}{(N-1)!}\frac{d^{N-1}}{dt^{N-1}}(t-t_{i})^{N}\phi_Z(t)e^{-itz}
\label{eq35}
\end{equation}
When $z>0$, the residue $r_{1}(z)$ is due to $t_{1}$. The residue
$r_{2}(z)$ is due to $t_{2}$ when $z<0$.

As an example, we consider the case when the number of the
transmit antennas is three, $N=3$. Evaluating the residues $r_1(z)$ and
$r_2(z)$ using (\ref{eq35}), we get
\begin{eqnarray}
r_{1}(z)&\!\!\!=\!\!\!&\frac{(\lambda\mu)^{3}}{2}\left(
\frac{z^{2}e^{-\mu z}}{(\lambda+\mu)^{3}}+ \frac{6ze^{-\mu
z}}{(\lambda+\mu)^{4}}+ \frac{12e^{-\mu z}}{(\lambda+\mu)^{5}}
\right) \label{35}\\
r_{2}(z)&\!\!\!=\!\!\!&\frac{(\lambda\mu)^{3}}{2}\left(
\frac{z^{2}e^{\lambda z}}{(\lambda+\mu)^{3}}- \frac{6ze^{\lambda
z}}{(\lambda+\mu)^{4}}+ \frac{12e^{\lambda z}}{(\lambda+\mu)^{5}}
\right) \label{36}
\end{eqnarray}
Therefore, the PDF of $z$ is given by
\begin{equation}
f_{Z}(z)= \left\{%
\begin{array}{cc}
  \frac{(\lambda\mu)^{3} e^{-\mu z}}{2(\lambda+\mu)^{3}}  \left(
z^{2}+ \frac{6z}{(\lambda+\mu)}+ \frac{12}{(\lambda+\mu)^{2}}
\right) & z > 0\\
  \frac{(\lambda\mu)^{3}e^{\lambda z}}{2(\lambda+\mu)^{3}}\left(z^{2}- \frac{6z}{(\lambda+\mu)}+ \frac{12e^{\lambda
z}}{(\lambda+\mu)^{2}} \right) & z < 0 \\
\end{array}%
\right. \label{37}
\end{equation}
A closed-form expression of the outage probability of the $i$th
user can be obtained then as
\begin{equation}
    P_{\text{out},i}=\int_{-\infty}^{\frac{N\sigma^{2}\gamma}{P}}f_{z}(z)\, dz
\end{equation}

In direct ARQ systems, if an outage event occurs at any user, the
user transmits a NACK signal to the BS. The BS retransmits the data vector again to the user.
Assuming that the channel coefficients are independent from one
transmission to another, the outage probability after $L$
retransmissions is:
\begin{equation}
    P^{\text{(ARQ,L)}}_{\text{out},i}=\left(P_{\text{out},i}\right)^L
\end{equation}

As an illustrative example, let us consider a 2-cell MISO system
with $N=3$ antennas at the BSs. We use
$\sigma^{2}=10^{-3}$, $\sigma_{1}^{2}=2$, $\sigma_{2}^{2}=1$, and
$R=2$ b/s/Hz. We define the transmit signal-to-noise ratio (SNR) as $SNR=\frac{P}{\sigma^{2}}$.
Fig.~\ref{fig_ARQ_L} shows the outage probability of
the system in the absence and presence of interference for
different number of retransmission attempts.
Even with ten retransmissions, the outage probability
of the system in presence of inter-cell interference is inferior to
that of a single user system. This can be attributed to the fact
that direct retransmission can overcome the fading, however, it
cannot overcome the interference.
\vspace*{-30mm}
\begin{figure}[ht!]
\centering
\includegraphics[width=90mm]{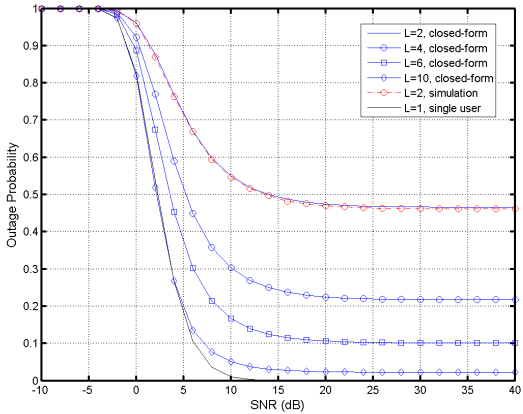}
\vspace*{-30mm}
\caption{Outage probability in the presence and absence of
interference for a direct ARQ system.} \label{fig_ARQ_L}
\end{figure}

\section{shared relaying for MISO ARQ systems}

In order to manage the inter-cell interference, we assume that a
MIMO half-duplex DF shared relay with $M$ antennas
is deployed to retransmit the lost packets to the target users. We
assume that the relay can successfully decode the messages
transmitted from the two BSs and the NACK signals
transmitted by the users. This assumption is well justified, since
the relay can be placed such that it has line of sight
communication with the BSs. We also assume that the
relay can acquire the transmit channel state information (CSI) to the
users at the cell edge. This can be achieved with much smaller
training overhead than that required to transmit the CSI to the BSs as the relay is placed in close
proximity to the cell edge users. Nevertheless, we assume that the
relay does not have any information about the channels from the
BSs to the users.

The relay operates in one of two modes. The first mode is the
single-user retransmission where only one user has failed to
receive its message. In this case, the relay retransmits the
signal for this user while the BS associated with the
other user {\it simultaneously} transmits a new message to its designated user. In
contrast, in the second mode, both users have failed to receive
their messages, and hence, the relay forwards these messages to
the users while the BSs are silent. In each case, we
present a relay beamforming algorithm that aims to improve the
outage probability of the system by managing the inter-cell interference associated
with downlink transmission.

\subsection{Single-User Retransmission}
Let us assume that the first user has decoded its message while
the other user has failed to decode its message. In the next time
slot, the first BS retransmits a new message to its
target user while the relay simultaneously retransmits the lost message to the
second user. The received signal by the $i$th user in this case is
given by
 \begin{equation}
{{y}}_{i}=\sqrt{\frac{P}{N}}{\boldsymbol{h}}_{i,1}^H{\boldsymbol{x}}_{1}
+{\boldsymbol{h}}_{i,r}^H{\boldsymbol{B}}{\boldsymbol{x}}_{r}
+{\boldsymbol{n}}_{i} \label{eq3}
\end{equation}
where ${\boldsymbol{x}}_{r}$ is the retransmitted message from the
relay and is equal to the decoded message of the second user in
the previous time slot, the $M\times N$ matrix ${\boldsymbol{B}}$
is the relay beamforming matrix, ${\boldsymbol{h}}_{i,r}$ is the
$M\times 1$ vector containing the complex conjugate of the channel
coefficients between the relay and the $i$th user. These channel
coefficients are independent identically distributed zero-mean
Gaussian random variables with variance $\sigma_3^2$.

In order to decrease the outage probability for the first user, we design the relay beamforming matrix
such that the relay transmission does not cause any interference.
In order to improve the outage probability of the second user, we
maximize the received signal power at the second user given the
constraint on the relay transmission power. Hence, we can write
the relay design problem as
\begin{equation}
   \max_{\boldsymbol{B}}\Big\|{\boldsymbol{B}}^H
   {\boldsymbol{h}}_{2,r}\Big\|^2
   \quad \text{s.t. }
   {\boldsymbol{B}}^H{\boldsymbol{h}}_{1,r}\!=\!\boldsymbol{0}_N,\;
   \text{tr}\left\{\!{\boldsymbol{B}}{\boldsymbol{B}}^{H}\!\right\}\!=\!P_{r} \quad\label{eq10}
\end{equation}
where $\text{tr}\{\cdot\}$ denotes the trace of a matrix and $P_r$
is the relay transmission power. For the sake of fairness of the
comparison with direct ARQ case, we use $P_r=P$. Using the
identity $\text{vec}\left\{{\boldsymbol{A}}{\boldsymbol{B}}
{\boldsymbol{C}}\right\}=\left({\boldsymbol{C}}^{T}
\otimes{\boldsymbol{A}}\right)
\text{vec}\left\{{\boldsymbol{B}}\right\}$ where
$\text{vec}\left\{\cdot\right\}$ is the vectorization operator and
$\otimes$ is the matrix Kronecker product. The optimization
problem in (\ref{eq10}) can be expressed as
\begin{eqnarray}
 \max_{\tilde{\boldsymbol{b}}} &\!\!\!\!\!\!& \tilde{\boldsymbol{b}}^H    \left( \boldsymbol{I}_N \otimes {\boldsymbol{h}}_{2,r}{\boldsymbol{h}}_{2,r}^H \right)
 \tilde{\boldsymbol{b}}\nonumber\\
   \qquad \text{s.t. }&\!\!\!\!\!\!& \left( \boldsymbol{I}_N \otimes
   {\boldsymbol{h}}_{1,r}^H\right) \tilde{\boldsymbol{b}} =
   \boldsymbol{0}_{MN},\, \big\|\tilde{\boldsymbol{b}}\big\|^2=P_{r}\label{eq11}
\end{eqnarray}
where $\tilde{{\boldsymbol{b}}}=
\text{vec}\left\{{\boldsymbol{B}}\right\}$. Let us define the
orthonormal  $MN\times (M-1)N$ matrix $\boldsymbol{V}$ such that
its columns span the null space of the matrix $\boldsymbol{I}_N
\otimes {\boldsymbol{h}}_{1,r}^H$. Therefore, we can write the
optimal solution of the optimization problem in (\ref{eq11}) as
\begin{equation}
\tilde{{\boldsymbol{b}}}= \sqrt{P_{r}}\; \boldsymbol{V}
\nu_{\text{max}}\left\{ \boldsymbol{V}^H \left(\boldsymbol{I}_N
\otimes {\boldsymbol{h}}_{2,r}{\boldsymbol{h}}_{2,r}^H \right)
\boldsymbol{V}\right\} \label{eq12}
\end{equation}
where $\nu_{\text{max}}\{\boldsymbol{A}\}$ is the  eigen vector of
the matrix $\boldsymbol{A}$ associated with its maximum eigen
value.

Similarly, we can calculate the mutual information between the
transmitting and receiving nodes as
\begin{eqnarray}
  {I}_{1}^{\text{(I)}} &\!\!\!=\!\!\!& \log_{2} \left(1+ \frac{
\frac{P}{N}{{\boldsymbol{h}}_{1,1}^H
{\boldsymbol{h}}_{1,1}}}{{\boldsymbol{h}}_{1,r}^{H}\boldsymbol{B}\boldsymbol{B}^H{\boldsymbol{h}}_{1,r}+\sigma^{2}}
\right)\label{eq4}\\
  {I}_{2}^{\text{(I)}} &\!\!\!=\!\!\!& {\log_{2}}\left(1+
  \frac{{\boldsymbol{h}}_{2,r}^{H}\boldsymbol{B}\boldsymbol{B}^H{\boldsymbol{h}}_{2,r}}{\frac{P}{N}{{\boldsymbol{h}}_{2,1}^H
  {\boldsymbol{h}}_{2,1}+\sigma^{2}}}\right)
\end{eqnarray}
t

\subsection{Multiuser Retransmission}
In the case when both users fail to decode their transmitted
messages, the relay forwards them to the users in the next time
slot while the BSs remain silent. The transmitted signal
by the relay in this scenario is given by
\begin{equation}
    \boldsymbol{x}_r=\boldsymbol{B}_1 \boldsymbol{x}_1+ \boldsymbol{B}_2
    \boldsymbol{x}_2
\end{equation}
The received signal at the $i$th user is given by
\begin{equation}
{{y}}_{i}= {\boldsymbol{h}}_{i,r}^H \boldsymbol{B}_1
\boldsymbol{x}_1+ {\boldsymbol{h}}_{i,r}^H \boldsymbol{B}_2
\boldsymbol{x}_2 +{\boldsymbol{n}}_{i} \label{eq6}
\end{equation}
As a result, we can write the received SINR of the $i$th user as
\begin{equation}
\text{SINR}_i=\frac{{\boldsymbol{h}}_{i,r}^H \boldsymbol{B}_i\boldsymbol{B}_i^H{\boldsymbol{h}}_{i,r}}
{{\boldsymbol{h}}_{i,r}^H \boldsymbol{B}_j\boldsymbol{B}_j^H{\boldsymbol{h}}_{i,r}+\sigma^2}
\label{SINR}
\end{equation}
where $i \in \lbrace 1,2 \rbrace$ and $i \neq j$.\\

In order to decrease the outage probability, we design the relay
beamforming matrices such that the minimum SINR of the two users
is maximized under the constraint on the relay transmission power.
Using the auxiliary variable $t$, we can write the relay design
problem as
\begin{eqnarray}
  \max_{\boldsymbol{B}_1,\boldsymbol{B}_2,t} &\!\!\!\!& t \nonumber\\
  \text{s.t. }&\!\!\!\!& \text{SINR}_{i} \geq t \qquad i=1,2
  \nonumber\\
  &\!\!\!\!& \text{tr}\left(\boldsymbol{B}_1
\boldsymbol{B}_1^H+ \boldsymbol{B}_2 \boldsymbol{B}_2^H\right)\leq
P_{r}\label{eq27}
\end{eqnarray}
where $P_r$ is chosen in this case as $P_r=2P$ for the sake of
fairness when comparing with the direct ARQ system. Problem (\ref{eq27}) can be solved by using the Bisection method
as a sequence of feasibility problems as follows \cite{boyd}. We
initialize the lower bound $b_l$ and the upper bound $b_u$ on the
objective function respectively as, $b_l=0$, and
\begin{equation}
    b_u=\min_{i=1,2} \frac{P_r{\boldsymbol{h}}^H_{i,r}{\boldsymbol{h}}_{i,r}}{\sigma^2}
\end{equation}
At each iteration, we solve a feasibility problem in the variables
$\boldsymbol{B}_1$ and $\boldsymbol{B}_2$ at $t=(b_l+b_u)/2$ for
the constraints in (\ref{eq27}). If the problem is feasible we set
$b_l=t$, else, we set $b_u=t$. The procedure is repeated until
$b_u-b_l \leq \epsilon$ where $\epsilon$ is the required accuracy
for the solution. The resulting feasibility problem is not convex due to the SINR
constraints in (\ref{eq27}). Let
$\tilde{\boldsymbol{b}}_i=\text{vec}\{\boldsymbol{B}_i\}$, then we
can write the feasibility problem as
\begin{eqnarray}
  \text{Find} &\!\!\!\!\!\!& \tilde{\boldsymbol{b}}_1,\tilde{\boldsymbol{b}}_2 \nonumber\\
  \text{s.t. } &\!\!\!\!\!\!& \text{tr}\! \left\{\boldsymbol{C}_i
 \tilde{\boldsymbol{b}}_i\tilde{\boldsymbol{b}}_i^H \!\right\} \geq t
 \text{tr}\! \left\{ \boldsymbol{C}_i \tilde{\boldsymbol{b}}_j\tilde{\boldsymbol{b}}_j^H \!\right\}+
 t\sigma^2 \quad i,j\!=\!1,2, i\!\neq\! j\nonumber\\
 &\!\!\!\!\!\!& \text{tr}\! \left\{
 \tilde{\boldsymbol{b}}_1\tilde{\boldsymbol{b}}_1^H \!\right\}+ \text{tr}\! \left\{
 \tilde{\boldsymbol{b}}_2\tilde{\boldsymbol{b}}_2^H \!\right\}\leq
 P_{r}\label{eq29}
\end{eqnarray}
where $ \boldsymbol{C}_i= \boldsymbol{I}_N \otimes
{\boldsymbol{h}}_{i,r}{\boldsymbol{h}}_{i,r}^H $ and
$\text{tr}\{\cdot\}$ denotes the trace of a matrix. Let us define
the $MN\times MN$ matrices $\boldsymbol{X}_i=
\tilde{\boldsymbol{b}}_i\tilde{\boldsymbol{b}}_i^H$.  We use
SDR to convert the problem in (\ref{eq29})
into a convex optimization problem. The relaxed problem can be
expressed as the following semi-definite program
\begin{eqnarray}
  \text{Find} &\!\!\!\!\!\!& \boldsymbol{X}_1,\boldsymbol{X}_2 \nonumber\\
  \text{s.t. } &\!\!\!\!\!\!& \text{tr}\! \left\{\boldsymbol{C}_i
 \boldsymbol{X}_i\!\right\} \geq t
 \text{tr}\! \left\{ \boldsymbol{C}_i \boldsymbol{X}_j \!\right\}+
 t\sigma^2 \quad i,j\!=\!1,2, i\!\neq\! j\nonumber\\
 &\!\!\!\!\!\!& \text{tr}\! \left\{
 \boldsymbol{X}_1 \!\right\}+ \text{tr}\! \left\{
 \boldsymbol{X}_2 \!\right\}\leq
 P_{r}\nonumber\\
 &\!\!\!\!\!\!& \boldsymbol{X}_i\succeq 0 \qquad i=1,2  \label{eq30}
\end{eqnarray}
where we have relaxed the problem by dropping the rank one
constraints on the matrices $\boldsymbol{X}_1$ and
$\boldsymbol{X}_2$.

The optimization problem (\ref{eq30}) is similar to problem (15)
in \cite{rank_cons} and has an arbitrary rank profile.
Nevertheless, Theorem 3.2 in \cite{rank_cons} states that, we can
generate another optimal solution $\left(\boldsymbol{Z}_1^\star,
\boldsymbol{Z}_2^\star \right)$ from the optimal solution of
(\ref{eq30}), $\left(\boldsymbol{X}_1^\star,
\boldsymbol{X}_2^\star \right)$, with a constrained rank profile
that satisfies
\begin{equation}
2 \leq \text{rank}^{2}\left\{\boldsymbol{Z}_{1}^{\star}\right\} +
\text{rank}^{2}\left\{\boldsymbol{Z}_{2}^{\star}\right\} \leq 3
\label{eq22}
\end{equation}
Since the summation at (\ref{eq22}) includes $2$ matrices and the
rank of each matrix is greater than zero, this implies that the
rank of $\left(\boldsymbol{Z}_n^{\star}\right)=1$, for all
$n=1,2$.

In the following, we present Algorithm1 \cite{rank_cons} to generate a rank one optimal
solution, $\left\{\boldsymbol{Z}_{i}^{\star}\right\}_{i=1}^2$,
from the arbitrary rank solution of (\ref{eq30}),
$\left\{\boldsymbol{X}_{i}^{\star}\right\}_{i=1}^2$, without any
loss of optimality. Let
$r_{n}=\text{rank}\left(\boldsymbol{X}_n^\star\right)$ and
$W=r_{1}^{2}+r_{2}^{2}$.\\
$\textbf{\text{while}}$ $W>3$ do
\begin{enumerate}
\item Decompose
$\boldsymbol{X}_n^\star=\boldsymbol{V}_n\boldsymbol{V}_n^H$,
$n=1,2$. \item Find  $\left(\boldsymbol\Delta_1,
\boldsymbol\Delta_2 \right)$ a nonzero solution of the following
system of linear equations:
\begin{eqnarray}
\text{tr}\! \left\{
 \boldsymbol{V}_1^H\ \boldsymbol{C}_1 \boldsymbol{V}_1 \boldsymbol\Delta_1 \!\right\}-t \;\text{tr}\! \left\{\boldsymbol{V}_2^H \left(\boldsymbol{C}_1 \right)\boldsymbol{V}_2 \boldsymbol\Delta_2 \!\right\}&\!\!\!\!=\!\!\!\!&0\\
  \text{tr}\! \left\{\boldsymbol{V}_2^H \boldsymbol{C}_2 \boldsymbol{V}_2 \boldsymbol\Delta_2 \!\right\}-t \;\text{tr}\! \left\{
 \boldsymbol{V}_1^H\ \left(\boldsymbol{C}_2 \right) \boldsymbol{V}_1 \boldsymbol\Delta_1 \!\right\}&\!\!\!\!=\!\!\!\!&0\\
  \text{tr}\! \left\{\boldsymbol{V}_1^H \boldsymbol{V}_1 \boldsymbol\Delta_1 \!\right\}+ \text{tr}\! \left\{
 \boldsymbol{V}_2^H\  \boldsymbol{V}_2 \boldsymbol\Delta_2
 \!\right\}&\!\!\!\!=\!\!\!\!&0\qquad
\label{eq35}
\end{eqnarray}
where $\boldsymbol\Delta_n$ is $r_n\times r_n$ Hermitian matrix
for all $n$; \item Find the eigenvalues
$\delta_{n,1},\ldots,\delta_{n,r_n}$ of $\boldsymbol\Delta_n$ for
$n=1,2$;
\item Determine $n_{0}$ and $k_{0}$ such that\\
$\vert\delta_{n_{0},k_{0}}\vert=\text{max}
\left\lbrace\vert\delta_{n,k}\vert:1\leqslant k\leqslant r_n,
1\leqslant n\leqslant2 \right\rbrace $. \item Compute
$\boldsymbol{Z}_n^\star\!\!=\!\! \boldsymbol{V}_n\left(\!
\boldsymbol{I}_{r_n}\!-\!\frac{1}{\delta_{n_{0},k_{0}}}\boldsymbol\Delta_n
\!\right)\boldsymbol{V}_n^{H}$ for $n=1,2$. \item Evaluate
$r_{n}=\text{rank}\left(\boldsymbol{Z}_n^\star\right)$ for $n=1,2$
and $W=\sum\limits_{n=1}^2r_{n}^{2}$;
\end{enumerate}
$\textbf{\text{end while}}$\\%
%
\section{simulation and analytical results}
In this section, we present simulation results that compare the
performance of the proposed relaying scheme with direct ARQ
systems. The outage probability is used as the performance metric.
The comparison is performed under equal power allocation between
the different systems. We assume only one retransmission round. In
simulations, we use $\sigma^{2}=1$, $\sigma_{1}^{2}=2$,
$\sigma_{2}^{2}=1$ and $\sigma_{3}^{2}=4$, and $N=3$.\\

Fig. 2 shows the performance of the proposed ARQ system with a
MIMO shared relay with $M=3$ and the direct ARQ system at
different transmission rates.
 The single user case serves as a reference for outage probability with no interference.
The proposed ARQ relay system works efficiently at high
transmission rates compared with the direct ARQ system even at
very high transmission rates.
\vspace*{-30mm}
\begin{figure}[ht!]
\centering
\includegraphics[width=90mm]{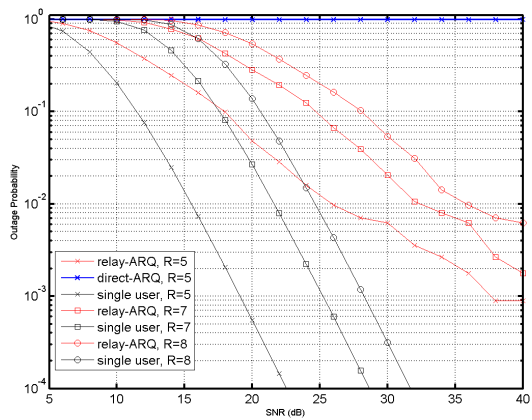}
\vspace*{-30mm}
\caption{Outage probability of the proposed ARQ relay system, direct ARQ and
 single user at different transmission rates .} \label{fig_ARQ_L}
\end{figure}\\

Fig. 3 depicts the performance of the proposed ARQ system with the
MIMO shared relay for different numbers of relay antennas at
certain transmission rate $R=6$. With sufficient number of
antennas at the relay, the proposed system model attains high
diversity gain, minimizes the interference, and achieves
performance even better than the single user case.

\begin{figure}[ht!]
\centering
\vspace*{-30mm}
\includegraphics[width=90mm]{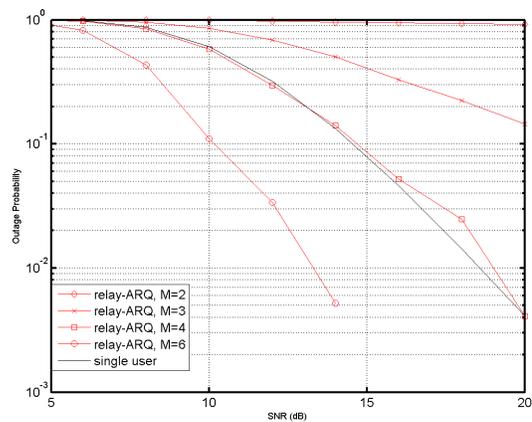}
\vspace*{-30mm}
\caption{Outage probability of the proposed ARQ relay system at different number
of relay antennas and single user case under R=6.} \label{fig_ARQ_L}
\end{figure}

\section{CONCLUSION}
In this paper, we have considered a MIMO shared relay operating in
the downlink of an ARQ wireless cellular system. We have proposed
relay beamforming techniques that improve the outage probability
by maximizing the received SINR at the users under a constraint on
the relay transmission power. The performance of the proposed
algorithms was compared to that of direct ARQ systems via
numerical simulations showing that significant performance
improvement can be achieved by the proposed system.

\vspace*{5mm}
\bibliographystyle{IEEEbib}
\bibliography{IEEEabrv,refrences2}
\end{document}